\documentclass[aps,prl,twocolumn,superscriptaddress,amsmath,amssymb,nofootinbib]{revtex4}
\usepackage[utf8]{inputenc}
\usepackage{graphicx}
\usepackage{dcolumn}
\usepackage{bm}
\usepackage{epsfig}
\usepackage{float}
\usepackage{booktabs}
\usepackage{amsmath}

\begin{document}

\title{In-situ measurement of the velocity spectrum of ultracold neutrons and its evolution using an oscillating detector} 

\author{D.~Rozpedzik}
\email[Corresponding author: ]{dagmara.rozpedzik@uj.edu.pl}
\affiliation{M. Smoluchowski Institute of Physics, Jagiellonian University, PL-30059 Cracow, Poland}
\author{K.~Bodek}
\affiliation{M. Smoluchowski Institute of Physics, Jagiellonian University, PL-30059 Cracow, Poland}
\author{K.~Kirch}
\affiliation{Paul Scherrer Institute, CH-5232 Villigen, Switzerland}
\affiliation{ Institute for Particle Physics and Astrophysics, ETH Zurich, CH-8093 Zurich, Switzerland}
\author{B.~Lauss}
\affiliation{Paul Scherrer Institute, CH-5232 Villigen, Switzerland}
\author{K.~Lojek}
\affiliation{M. Smoluchowski Institute of Physics, Jagiellonian University, PL-30059 Cracow, Poland}
\author{I.~Rienaecker}
\affiliation{Paul Scherrer Institute, CH-5232 Villigen, Switzerland}
\author{P.~Schmidt-Wellenburg}
\affiliation{Paul Scherrer Institute, CH-5232 Villigen, Switzerland}
\author{G.~Zsigmond}
\affiliation{Paul Scherrer Institute, CH-5232 Villigen, Switzerland}

\begin{abstract}
Ultracold neutrons (UCNs) are used in experiments investigating fundamental interactions, testing the Standard Model of particle physics and searching for phenomena beyond it. Knowledge of the energy spectrum of UCNs is very often a key ingredient to determine the systematic effects in precision measurements utilizing UCNs. The proposed novel method allows for the in-situ measurements of the UCN velocity distribution and its time evolution. In addition, the proposed UCN spectrometer can be a handy diagnostic tool for monitoring the spectrum in critical places in the UCN transport system connecting an UCN source with experiments. In this paper, we show the results from the measurements performed at the Paul Scherrer Institute's UCN source.
\end{abstract}

\date{\today}

\maketitle

Ultracold neutrons (UCNs) are free neutrons with velocities lower than about 8 m/s. They can be trapped for long times by specially formed magnetic fields or vessels coated with suitable materials having a high neutron optical potential (Fermi potential) \cite{1sears}. Trapped UCNs are excellent probes for investigating fundamental properties of the neutron. An additional experimental  advantage of using UCNs is the possibility of achieving fully polarized neutron samples and low background during measurements.

The UCN energy spectrum and its time evolution are often the key ingredients for assessment of systematic effects influencing the experimental uncertainty of measurements. This applies, especially, to high-precision experiments testing the Standard Model of particle physics and looking for physics beyond it. Among them are measurements of the neutron lifetime \cite{2gonzalez}, neutron beta-decay angular correlations \cite{3hasan,4brown,5perkeo}, search for the neutron electric dipole moment \cite{6abel},  searches for neutron-antineutron and neutron-mirror-neutron oscillations \cite{7fomin,8ayres,9berezhiani}, axion-like particles - dark matter candidates \cite{10afach,11abel} and many others \cite{12dubbers}. Most of these experiments are statistically limited by the number of available UCNs. The UCN density provided to experiments, in turn, depends strongly on the properties of transport guides connecting UCN sources with experiments. A compact and versatile spectrometer for diagnosis would greatly help improving the optimization process of the UCN transport guides and reaching a high UCN intensity. 

Traditionally the UCN energy or velocity distribution in experiments was measured using dedicated techniques based on the selection of UCNs within a narrow band of kinetic energies detected with a known efficiency.
The preferred selection is made with a time-of-flight technique e.g. using mechanical choppers \cite{13fierlinger}. Since UCNs interact with materials via the neutron optical potential and are influenced by magnetic fields or gravitation, the UCN energy distribution can also be deduced from \cite{14golub,15ignatovitch,16amsler}: (i) transmission through thin foils made of materials with different Fermi potential barriers (acting as cut-offs), (ii) transmission through an absorbing gas at different pressures, (iii) transmission through magnetic field barriers of different strengths, (iv) UCN absorption at different heights above a reflective floor (v) use of rotatable 'inverted-U' shaped UCN guides oriented at different angles against gravity, and (vi) measurement of the range in the gravitational field for highly collimated beams. An extensive discussion of these methods can also be found in \cite{17knecht}. In these measurements, controlling the UCN detection efficiency as a function of kinetic energy is challenging and often a source of systematic bias. All the above mentioned methods are cumbersome and involve systematic effects that are energy dependent. In general these arise from reflection of UCNs in the guiding systems, as any reflection leads to an energy dependent loss probability. Their estimation unavoidably relies on Monte Carlo simulation which, in turn, depends on the correct description of material properties such as elemental and isotopic compositions, surface properties, time constants of aging effects and many others. Moreover, most of these methods are quite inefficient as they are based on selecting only a small portion of the available UCNs.  

Application of the spin-echo method developed at the Paul Scherrer Institute (PSI) \cite{18echo} solved the problem of the in-situ energy spectrum measurement for specific situations when: (i) UCNs are polarized and (ii) the storage volume is embedded in a magnetic coil system such that the spin precession of an UCN ensemble can be manipulated with radio frequency signals. Unfortunately, this method is hardly applicable as a routine diagnostics and may be used only in special cases. 

A spectrometer based on the method applied in this article allows for a direct measurement of the UCN velocity component which is parallel to the oscillation direction of the UCN detector. The detection principle is based on detecting UCNs that overcome the Fermi potential of the detector surface layer. By moving the detection layer, a time-dependent effective Fermi potential is experienced by the UCNs. A major advantage of the oscillating spectrometer is that it can measure in any spatial direction in contrast to the gravitational selector where only the vertical UCN velocity component can be scanned. In contrast to a detector based on velocity selection via different Fermi potentials, which are only stepwise available, the oscillating detector allows for a quasi-continuous velocity detection. Moreover, the entire range of UCN velocities can in principle be covered in a single oscillation period, measuring the entire spectrum in identical conditions. Thus, it is possible to study the spectral evolution, a very desirable feature in the diagnostic of the UCN precision experiments where knowledge of the neutron velocity spectrum is important. 

This novel method is based on the following principles:
in first approximation, a UCN approaching a flat piece of the material surface experiences a repulsive or attractive short range force due to a Fermi potential defined as $V_{\mathrm F} = 2\pi {\hbar}^2 N b/m_n$, where $N$ denotes the number density of the material, $m_n$ is the neutron mass, and $b$ is the average neutron bound scattering length. For simplicity, one assumes that this surface has the size comparable with the de Broglie wavelength of the neutron and the potential barrier is constant over that surface, meaning that the force is perpendicular to that surface. In consequence, the interaction changes only the UCN momentum component perpendicular to the surface at the point of incidence. For a repulsive potential, an incident neutron will penetrate through the surface only, if its velocity component perpendicular to that surface is greater than the Fermi velocity $v^{\mathrm F}$ of the material, corresponding to the Fermi potential by the relation $v^{\mathrm F} = \sqrt{2V_{\mathrm F}/m_n}$. Quantum tunneling phenomena are neglected in this approximation. A detailed discussion on this approximation can be found e.g. in Ref.~\cite{19byrne}.

\begin{figure}[h]
\centering\includegraphics[width=1\columnwidth,height=0.27\textheight]{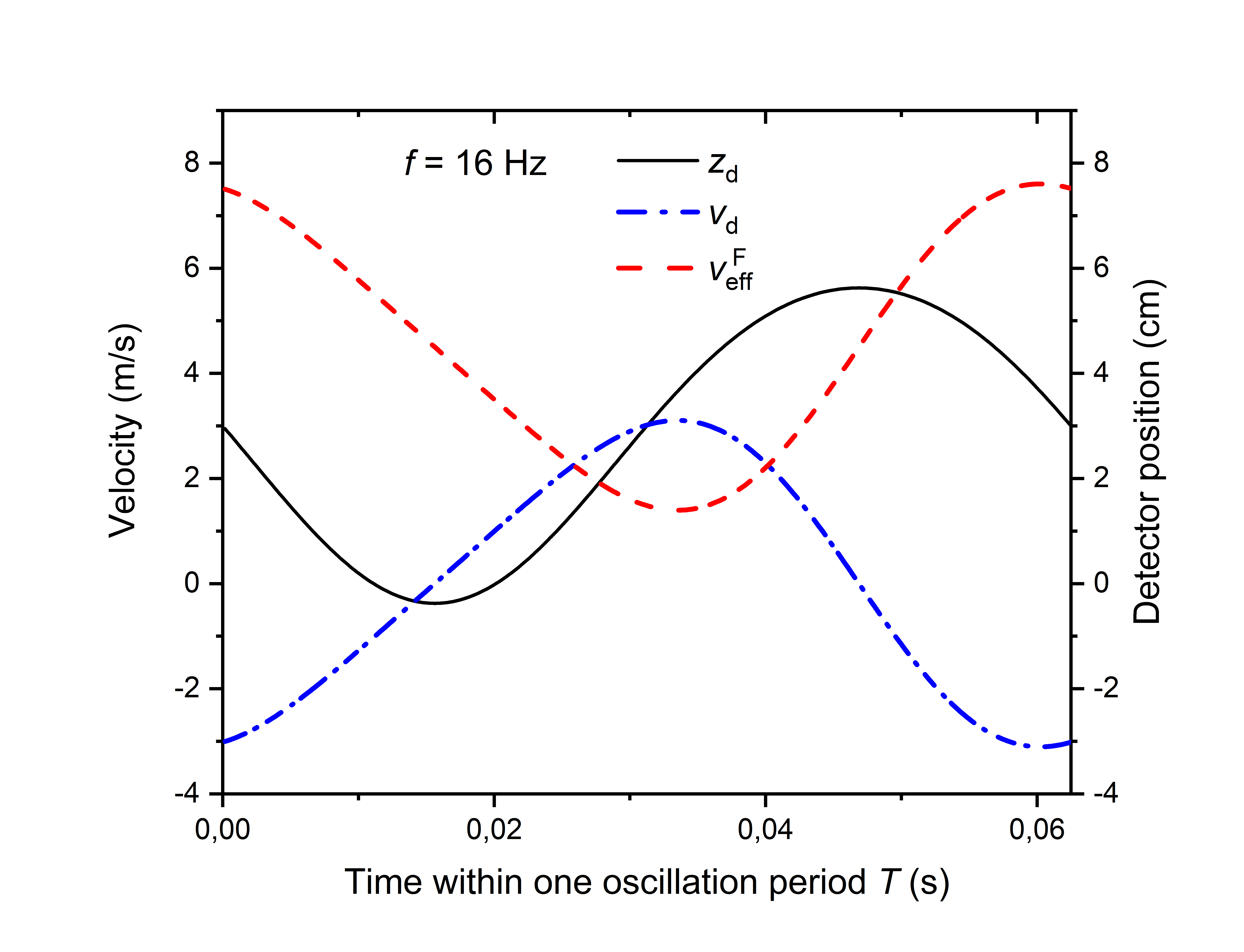} 
\caption{\label{fig-2}(color online) Time dependence (within one oscillation period) of the detector position $z_{\mathrm d}$ (black solid line) and detector velocity $v_{\mathrm d}$ (blue dash-dot line).
The red dashed line describes the Fermi velocity modulation of the detector surface for the example of a 16 Hz frequency.}
\end{figure}

The oscillating UCN spectrometer makes use of the mechanical acceleration of the detector towards incoming neutrons in order to change the neutron velocity as seen in the reference frame of the detector. At the moment of collision, the detector's active surface moves with the velocity $v_{\mathrm{d}}$, so that neutrons with a velocity component parallel to the detector motion $v_n > v^\mathrm{F}_{\mathrm d} + v_{\mathrm d}$ will penetrate into the material by passing the potential barrier of the detector and be detected. In other words, the moving detector changes the relative velocity of the neutron normal to the material surface. This may also be interpreted as a change of the Fermi potential of the detector surface. 
\begin{figure}[t]
\centering\includegraphics[width=1\columnwidth,height=0.3\textheight]{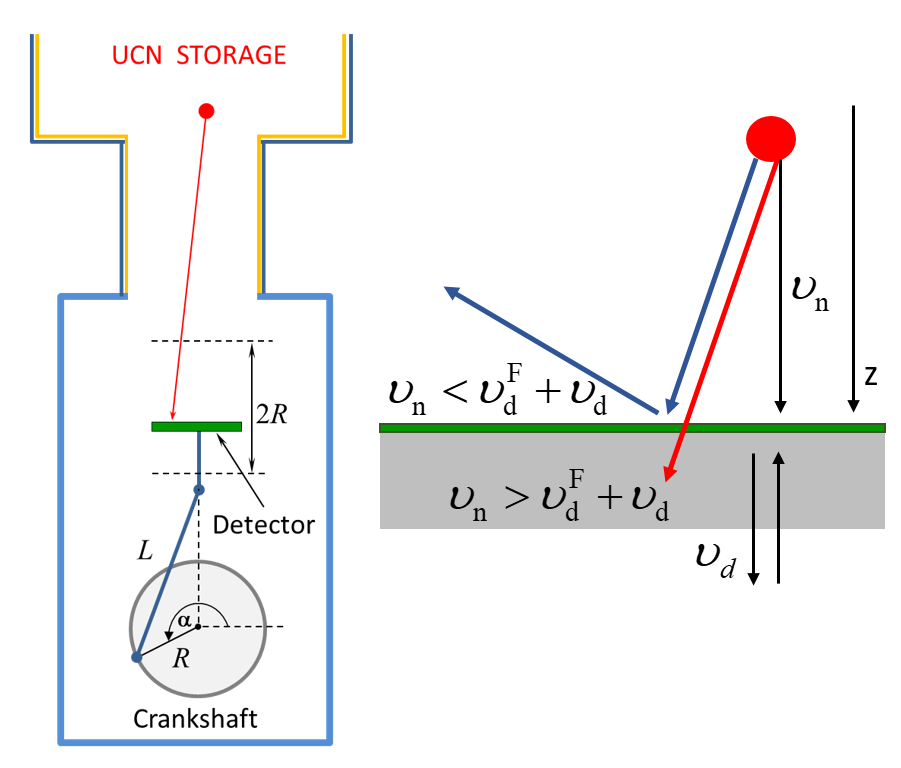}
\caption{\label{fig-1}(color online) General concept of the UCN oscillating spectrometer. Left panel: Crankshaft oscillator with $L=$ 0.12 m. The blue line marks the UCN absorbing aluminium walls of the chamber with Fermi velocity $v^{\mathrm F}_{\mathrm {Al}}$= 3.2 m/s. The yellow line identifies the Ni-Mo coated layer of UCN guide with Fermi velocity ($v^{\mathrm F}_{\mathrm s}$= 6.3~m/s) typical for a storage vessel. The green box denotes the detector surface with Fermi velocity $v^{\mathrm F}_{\mathrm d}$= 4.5 m/s attached to the crankshaft at angle  $\alpha= 2\pi ft$. Right panel: An incident neutron will penetrate through the moving barrier only, if its velocity component perpendicular to the surface, $v_n$, is greater than the sum of Fermi velocity of the detector surface and the detector velocity $v_\mathrm{d}$. The moving detector changes the relative perpendicular velocity of the neutron which may be interpreted as a change of the Fermi potential.
}
\end{figure} 
The spectrometer geometry is sketched in Fig.~\ref{fig-1}. The detector attached to a mechanical crankshaft is driven by a rotating axis. For the constant rotation frequency $f$ the detector position $z_{\mathrm d}(t)$ can be expressed by
\begin{equation}
z_{\mathrm d}(t) = R\left[1 - \sin (2\pi ft) - \frac{R}{4L} \left(1 - \cos (4\pi ft)\right)\right],
\label{eq-2}
\end{equation}
where $R$ is the crankshaft radius corresponding to the amplitude of the detector motion and $L$ describes the length of the connecting rod. 
The detector velocity $v_{\mathrm d}(t)$ is given by the equation
\begin{eqnarray}
v_{\mathrm d}(t) = 2\pi fR \left[-\cos (2\pi ft) + \frac{R}{2L} \sin (4\pi ft)\right].
\label{eq-3}
\end{eqnarray}
The time dependence of the detector position, detector velocity, and the Fermi potential modulation of the detector surface are illustrated in Fig.~\ref{fig-2}. 
If the detector oscillates in a way that the appropriate range of relative velocities is covered, all UCNs with a velocity component along the oscillation direction in a given range get a chance to be detected. Simultaneously, the acquisition system stores the time and, consequently, the detector velocity $v_{\mathrm d}$ of the event. In this way, the effective Fermi velocity at the instant of registration can be determined individually for each event  and related to the minimum UCN velocity necessary for crossing the detector potential barrier. Sorting the events according to the detection time (within the oscillation period $T$) and transforming the distribution will result in the UCN velocity spectrum. 

\begin{figure}[]
\centering\includegraphics[width=1.00\columnwidth,height=0.24\textheight]{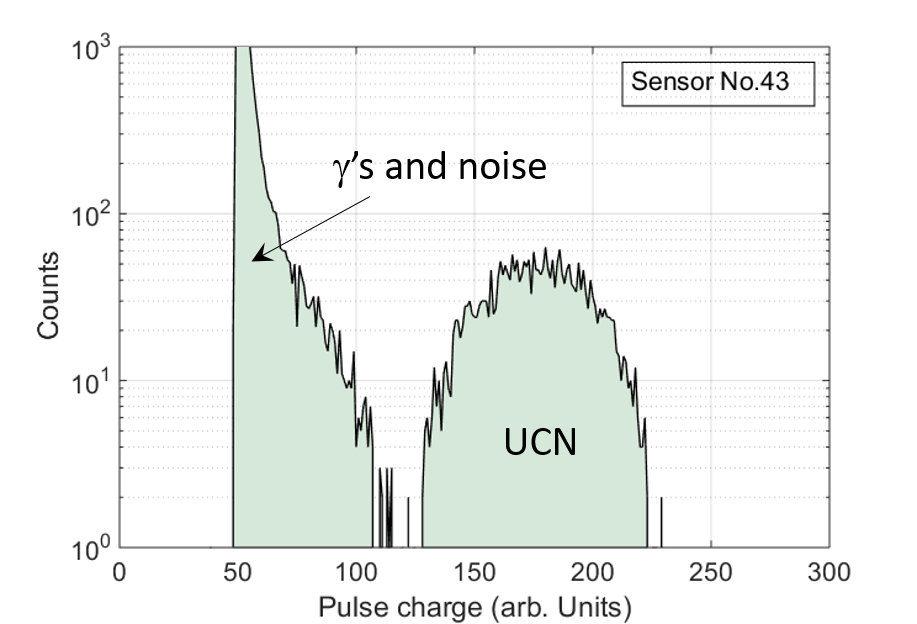} 
\caption{\label{fig-3}(color online) Pulse height spectrum of signals delivered by one sensor from the MPPC array attached to a 200 $\mu$m thick GS20 Li-glass scintillator exposed to UCNs.}
\end{figure}

In the present experiment the UCN detection is based on the neutron capture reaction on the ${^6}$Li isotope:
\begin{equation}
n + {^6}\mathrm{Li} \rightarrow {^3}\mathrm{H}\, (2.74\,\mathrm{MeV}) + {^4}\mathrm{He}\, (2.05\, \mathrm{MeV}).
\label{eq-1}
\end{equation}
For this purpose a 200 $\mu$m thick scintillator GS20 \cite{20gs20} with the Fermi velocity $v^{\mathrm F}_{\mathrm d}$ = 4.5 m/s  was optically coupled to an array of multi-pixel photon counter sensors (MPPC - S13361 series) \cite{21mppc}. Lithium-glass scintillators were successfully used for UCN detection e.g.~in Refs.~\cite{22gotl, 23ban,24rozpedzik}. The small thickness of the scintillator reduces the sensitivity to thermal neutrons and decreases $\gamma$-ray induced background. Additionally, $\gamma$-ray counts can be rejected by the Pulse-Shape Discrimination (PSD) method~\cite{23ban,24rozpedzik} since $\gamma$-ray signals do not show a slow decaying component and therefore are shorter than the scintillation signals generated by recoiling $\alpha$-particles and tritons in the Li-glass scintillator. From the collected charge distribution of the registered signals, it was found that electronic noise and gammas were well separated from the signal induced by neutron capture as can be seen in Fig.~\ref{fig-3}. Detailed information about the detector electronics used and the data treatment will be published in a seperate paper~\cite{25rozpedzik}.
In order to cover the UCN velocity range $0 - 8$ m/s in an oscillation period using a detector with a hypothetical $v^{\mathrm F}_{\mathrm d} = 0$, the product $ R\cdot f $ should be about 1 m/s. For the fixed oscillation amplitude $R$, the required UCN velocity range can be covered by adjusting the oscillation frequency $f$ and taking into account the offset introduced by the Fermi velocity of the detector material $v^{\mathrm F}_{\mathrm d}$. Given the amplitude $R=3$ cm and a maximum oscillation frequency $f = 16$ Hz attainable in the present mechanical system, and with $v^{\mathrm F}_{\mathrm d} = 4.5$ m/s, the detector velocity covers the range $v_{\mathrm d} \in (-3.02, 3.02)$ m/s and the scanned UCN velocity range of $1.48 - 7.52$ m/s. In order to shift the range towards higher velocity, the Fermi velocity of the detector surface needs to be increased by covering it with e.g. a Ni-Mo layer having $v^{\mathrm F}_{\mathrm d}$ = 6.3 m/s. The resulting accessible UCN velocity range would then become $3.48 - 9.02$ m/s. 

\begin{figure*}[]
\centering\includegraphics[width=2\columnwidth,height=0.38\textheight]{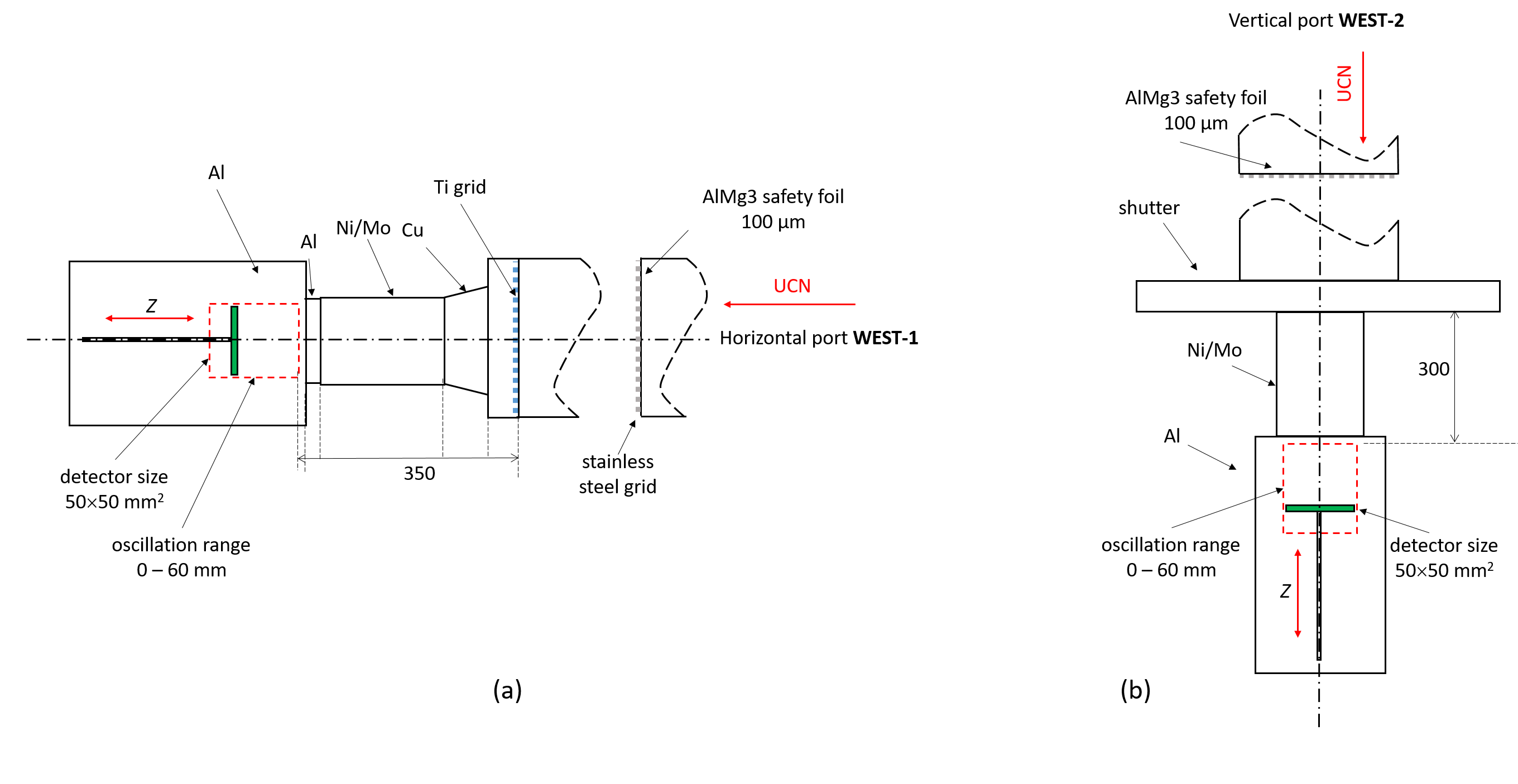} 
\caption{Measurement setups at PSI. On the left panel: the setup mounted on WEST-1 UCN beamport, on the right: the setup on WEST-2 beamport. All measures are given in mm.}
\label{fig-4}
\end{figure*}

\begin{figure*}[t]
\centering\includegraphics[width=2\columnwidth,height=0.38\textheight]{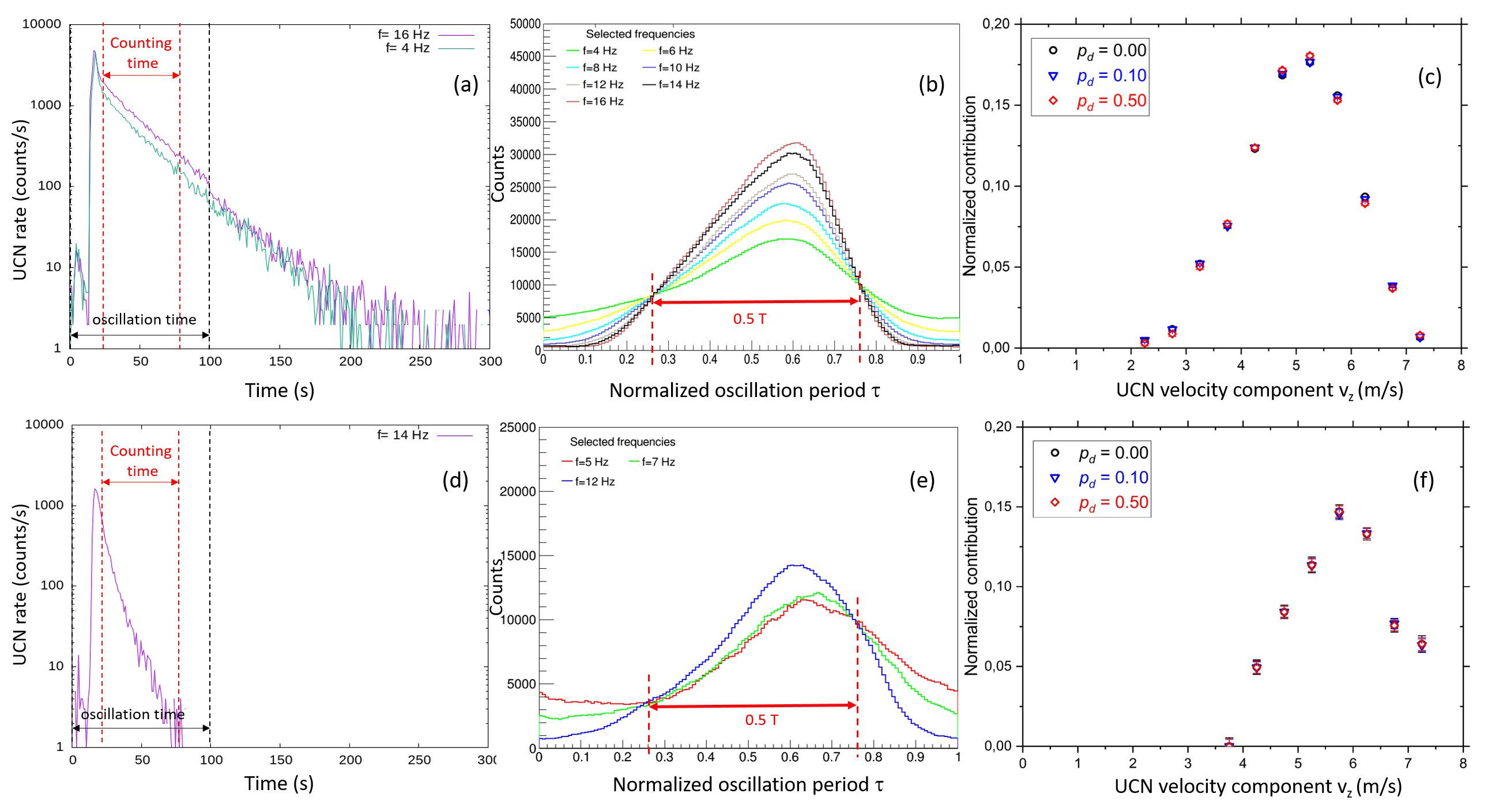}
\caption{\label{fig-5}From left (a) Total counting rate registered in a 300 s long beam-period at the WEST-1 beamport. The marked oscillation time corresponds to the time when the detector was moving. The counting time (area between the red dashed lines) is the time for which the data was analyzed. The small peak - 10 s before the main peak is caused by a 5 ms pilot proton beam pulse used for checking the beam centering before the main beam pulse on the spallation target \cite{26bison}. (b) Normalized time distributions at different frequencies for data collected within counting time set to 10 - 80 s after proton pulse. Distributions represent the running averages over 7 neighboring data points. (c) Velocity spectra corresponding to the time interval indicated in (a) obtained in the decomposition procedure for three values of the diffusion parameter $p_d$ in the MCUCN model. The different $p_d$ parameters reflect different assumptions for the detector entrance region. The bottom panels ((d),(e),(f)) display the same results obtained on the WEST-2 beamport. The counting time window was set to 10 - 80 s after proton pulse. Due to the extraction of UCNs from the top of the UCN source storage volume, the UCN intensity drops to almost zero after about 70 s, as the UCN energies in the storage vessel are to small to reach the height of the extraction guide \cite{26bison}.}
\end{figure*}

At the Paul Scherrer Institute's ultracold neutron source,  UCNs are produced and stored in a pulsed mode \cite{26bison,27lauss}. The UCN velocity spectra were measured at PSI using dedicated setups on WEST-1 and WEST-2 beamports as sketched in Fig.~\ref{fig-4}.
The UCN events in a defined acquisition time window were histogrammed within the oscillation period $T$ using a normalized quantity $\tau = \mathrm{mod}(t,T)/T$, where $t$ is the detection time given by the data acquisition system. Since the measurements at different oscillation frequencies were carried out independently, there was no guarantee that they correspond to identical flux of incoming UCN due to changes in UCN source intensity \cite{28anghel}. To account for this change in UCN intensity, the velocity need to be scaled to the integral flux measured with the detector at rest. During an oscillation period $T$ the detector comes twice to rest at the minimum and maximum position visible in Fig.~\ref{fig-2}. At these moments the measured flux is independent of the oscillation frequency neglecting a small effect caused by a slow drift of the energy distribution of the incoming UCNs (typical for the PSI source) within the short measurement time. The measured distributions should be scaled in order to cross at these points. Before scaling, the measured time profiles were smoothed using a weighted running average of the 7 neighboring data points in order to decrease statistical fluctuations. This step allows the adjustment of the scaling factors accurately and improves the stability of the decomposition procedure as described in the following. Fig.~\ref{fig-5} (panels (b) and (e)) shows the resulting normalized experimental time distributions for the two UCN exit beamports WEST-1 and WEST-2, respectively. 
In panels (a) and (d), the corresponding time intervals for event integration are indicated as 10 - 80 s after proton pulse.

On the one hand, the decomposition of the measured time profiles was performed under the assumption that the flux density $G^f(t)$ of the incident UCN changes slowly with respect to the oscillation period $T$ meaning that the rate of the detected UCN by a detector oscillating with the constant frequency $f = 1/T$ changes periodically in time. On the other hand, this flux density consists of contributions from UCNs belonging to different velocity bins, thus the total flux density of detected UCNs can be written as
\begin{eqnarray}
G^{f}(\tau_i) = \sum_{ij}F_{ij}^{f}(\tau_i,v_j^{\mathrm{UCN}})A_j(v_j^{\mathrm{UCN}}),
\label{eq-4}
\end{eqnarray}
where $\tau_i$ denotes a normalized time bin (within the oscillation period $T$), $v_j^{\mathrm{UCN}}$ is the $j$-th bin of the UCN velocity component along the oscillation direction measured in the laboratory reference frame, $A_j$ denotes the (unknown) UCN flux density contribution with velocity contained within $j$-th bin. $F_{ij}^{f}$ is the spectrometer response (or transmission) function. It connects the velocity bin $j$ with the detection time bin $i$. $F_{ij}^{f}$ can be defined as probability that a neutron with the velocity $v_j^{\mathrm{UCN}}$, entering randomly the volume swept by the oscillation detector within period $T$, will be registered in the time bin $t_i$. For the simple geometry of the setup this quantity can be calculated analytically. However, for our analysis, the response function $F_{ij}^{f}$ was obtained with the help of Monte Carlo simulations by including in MCUCN \cite{29geza} a detailed model of the oscillation of the detector surface. In this way, secondary effects such as diffusive scattering in the vicinity of the detector and kinetic cooling/heating of UCNs by the moving detector, and the flux modulation due to the oscillating relative velocity of the sensitive surface were more accurately taken into account.

The UCN velocity distribution $A_j(v_j^{\mathrm{UCN}})$ was obtained from measurement data by solving the linear pseudo-inverse problem with Tikhonov regularization~\cite{30engl,31calvetti}. 
Hence, the Eq.~\ref{eq-4} can be written in a matrix form as:
\begin{equation}
\mathbf{G = FA}, \quad F \in  \mathbb{R}^{m \times n},\quad G\in \mathbb{R}^{m}, \quad A\in \mathbb{R}^{n} 
\label{eq-5}
\end{equation}
Here, $n$ denotes the number of UCN velocity bins and $m$ is the number of time bins multiplied by the number of data sets collected at different frequencies $f$ that are included in the spectrum evaluation algorithm. The number of data sets necessary for a stable solution was chosen by testing the decomposition algorithm using synthetic data.
The optimum number of frequencies depends on the statistical quality of the  data. It turns out that in certain cases a single data set can be decomposed reliably, however, more stable solutions are obtained, typically, with three oscillation frequencies. The accuracy of time stamping and the rotation phase encoding assures that $m > n$, and therefore Eq.~(\ref{eq-5}) represents an overdetermined linear system which can be solved as a linear least--squares problem of minimal Euclidean norm of Eq.~(\ref{eq-5}):
\begin{equation}
\mathrm{min}_{A\in  \mathbb{R}^n}||FA - G||^{2}.
\label{eq-6}
\end{equation}
Furthermore, the system ~(\ref{eq-5}) can be ill-conditioned.

In order to get a stable solution one has to suppress effects of statistical noise in the measured distributions. This was achieved by applying a running average to the experimental distributions and adding the Tikhonov regularization which replaces the linear system of Eq.~(\ref{eq-5}) by the minimization problem:
\begin{equation}
\mathrm{min}_{A\in \mathbb{R} ^n} \{||FA - G||^2 + \lambda^2 ||LA||^2\},
\label{eq-7}
\end{equation}  
where $\lambda$ is the regularization parameter and $L$ is the regularization matrix, see e.g. \cite{32Lreg,30engl,25rozpedzik}. 
In further calculations the singular value decomposition (SVD) method was applied for $F$ and based on the discrete Picard plot a solution region in velocity bins was found that gives stable results \cite{30engl,25rozpedzik}. This region corresponds to the actual velocity range of the UCNs registered.

The value of $\lambda$ was chosen based on the L-curve method  by looking for the point of maximum curvature of the function \cite{30engl,25rozpedzik}:
\begin{equation}
\ell := \{(\log||LA_{\lambda,L}||,\log||FA_{\lambda,L}-G||):\lambda \geq 0\},
\end{equation}
where $||FA_{\lambda,L} -G||$ is the residual term called data fitting term and $||LA_{\lambda,L}||$ is the regularization term. The parameter $\lambda$ balances these two terms. Sometimes it happens that the region with the maximum curvature is not clearly visible. In such a case, the correctness of the choice of the regularization factor was checked by looking at the behavior of the residual and the relative error of solutions \cite{25rozpedzik}.

The statistical error of the resulting distribution $A_{\lambda,L}$ was estimated by propagating the covariance matrix of the experimental time distributions $G$ to the running average $\langle G \rangle$ and, finally, through the least squares solution of Eq.~(\ref{eq-7}) according to 
\begin{eqnarray}
\mathrm{Cov}(A_{\lambda,L})&=&{(F^TF +\lambda^2 L^TL)}^{-1}F^T\,\mathrm{Cov}(\langle G \rangle)\nonumber\\
&& F{(F^TF+\lambda^2 L^TL)}^{-1} .\nonumber\\
\label{eq-18}
\end{eqnarray}

The resulting UCN velocity spectra for WEST-1 and WEST-2 beamports integrated over indicted time intervals are plotted in panels (c) and (f) of Fig.~\ref{fig-5}, respectively.
\begin{figure*}[t]
\centering\includegraphics[width=\columnwidth,height=0.33\textheight]{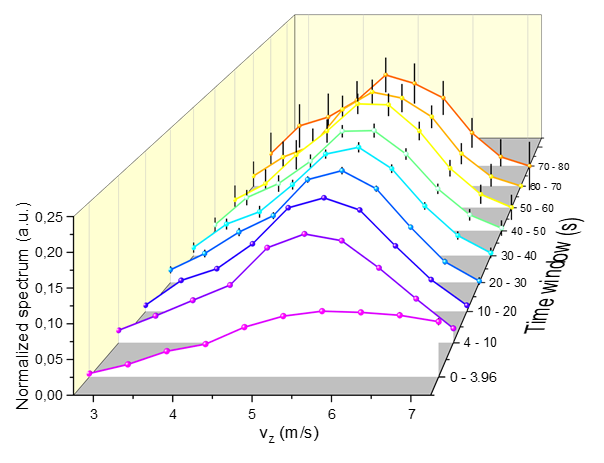}
\centering\includegraphics[width=\columnwidth,height=0.33\textheight]{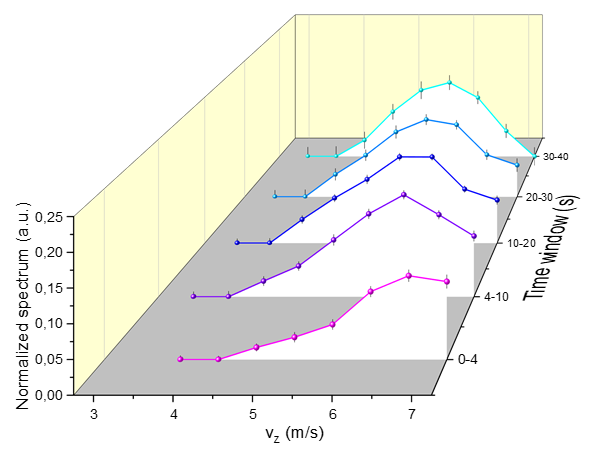}\\
\caption{(color online) Evolution of the $v_z$ component of the UCN velocity on WEST-1 beamport (left panel) and WEST-2 beamport (right panel).}
\label{fig-6}
\end{figure*}
\begin{figure*}[t]
\centering\includegraphics[width=\columnwidth,height=0.28\textheight]{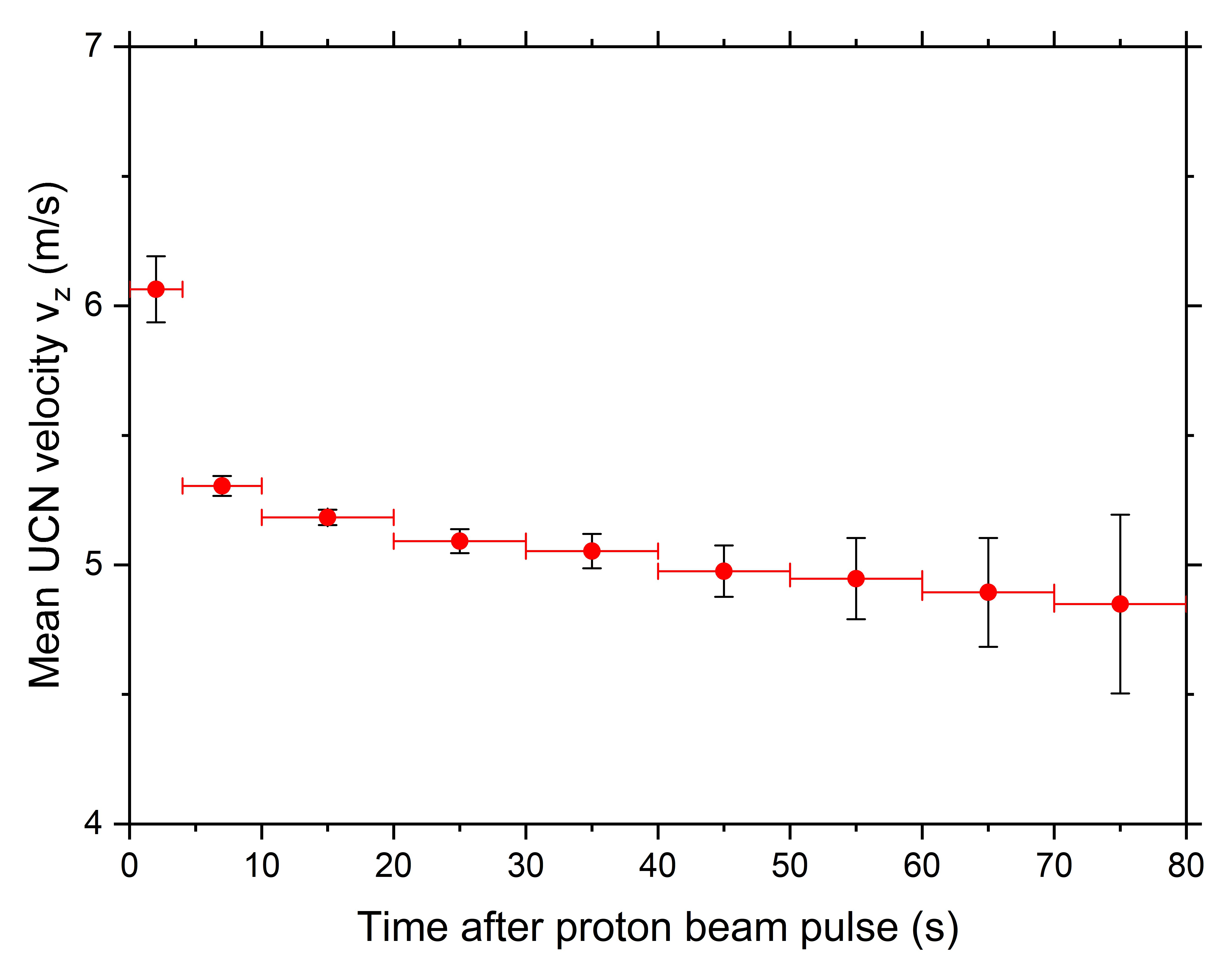}
\centering\includegraphics[width=\columnwidth,height=0.28\textheight]{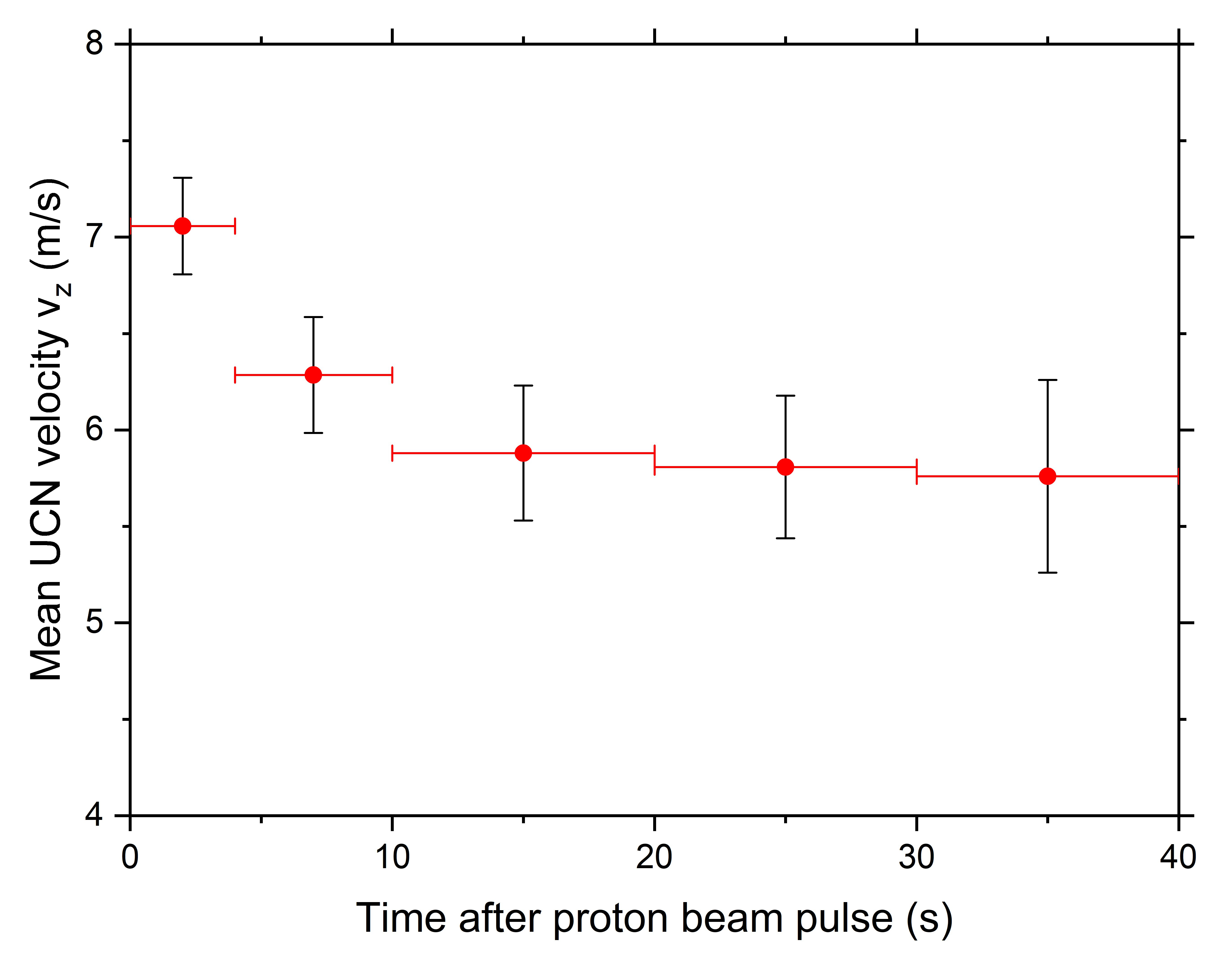}
\caption{(color online) Evolution of the mean value of the UCN velocity distribution measured at WEST-1 beamport (left panel) and
WEST-2 beamport (right panel). The scanned UCN velocity range was 1.5 - 7.5 m/s. Horizontal error bars indicate the period of time being investigated.}
\label{fig-7}
\end{figure*}

At the WEST-1 beamport the UCN energy spectrum starts at an energy of ~54 neV ($v_n$ = 3.25 m/s) due to the Fermi potential of the AlMg3 safety window mounted inside the neutron guide. At the WEST-2 beamport, the minimum UCN energy is ~130 neV ~\cite{26bison,33bison} because of the vertical drop of the UCN guide where UCNs are accelerated by gravity. It should be kept in mind that the oscillating spectrometer senses only the component of the UCN velocity which is parallel to the oscillation direction. On the way from the AlMg3 safety window, i.e. the place of definition of the minimum nominal velocity to the scanned volume some UCNs may undergo diffuse scattering causing UCN energies below the nominally possible value. This applies, especially, to measurements with longer storage times.

Stored UCN change their energy distribution in time due to collisions with storage walls. UCN can be lost due to small gaps in the storage vessel, surface imperfections (local spots of lower Fermi potential) and upscattered by phonon interactions during the wall collisions. The losses depend on the storage geometry, surface properties and initial UCN energy spectrum. It can be modelled by Monte Carlo simulation (see e.g.~\cite{29geza,33bison}) by assigning to the particular setup an effective parameter $\eta$ describing a probability of the UCN loss per single wall collision. 
UCN losses can be estimated from fits to the UCN storage curve (as e.g. in Figs.~\ref{fig-5}, panels: (a), (d), and in Refs.~\cite{26bison,33bison} ). 
In this way, the UCN spectrum evolution can be modelled. 
It should be stressed that the presented analysis of experimental data requires very weak model assumptions with the largest one being $p_d$, a parameter representing a fraction of the diffusive UCN scattering in the vicinity of the oscillating detector. $p_d$ is one of input parameters for the MCUCN calculation of the transmission function $F$ and gives dominating contribution to the systematic uncertainty of the extracted velocity distribution. This was estimated by performing the analysis with the transmission coefficients calculated for boundary values of $p_d \in (0.02,0.50)$ (see Fig.~\ref{fig-5}). The resulting differences are comparable with statistical errors. Another important systematic effect is due to reflections from the moving detector. These UCNs may reach again the detector at a later time. Monte Carlo calculations show that the effect of kinetic cooling/heating of the original spectrum is on the level of $10^{-3}$ or lower \cite{25rozpedzik} and thus negligible in the presented analysis.  

The presented method allows for empirical study of the UCN spectrum evolution. The spectrum decomposition can be confined to selected intervals of the UCN storage time. 
The width of these time intervals must be optimized to result in statistically significant profiles. The whole analysis process was repeated for subsequent storage-time intervals. The obtained evolution of the UCN velocity is illustrated in Fig.~\ref{fig-6} for the two UCN beamports WEST-1 and WEST-2, respectively. In order to better recognize the systematic softening of the spectrum, plots shown in Fig.~\ref{fig-7} represent the evolution of the mean UCN velocity $v_n$
within the storage time interval.

In this paper, we described an in-situ measurement of the UCN velocity distribution and its development during storage. In this context, "in situ" means a volume of $50 \times 50 \times 60$ mm$^3$ scanned by the oscillating detector. The measured velocity component is parallel to the oscillation direction. For the port WEST-1 the oscillating spectrometer was oriented horizontally while for the WEST-2 beamport the scanning direction was vertical. The velocity spectrum on WEST-1 can be compared to the results in Ref. \cite{34ingo}. In order to extract the full velocity distribution, additional measurements in orthogonal directions are planned. 
It should be stressed that the decomposition method based on solving the pseudoinverse problem of the linear system of equations applied to the experimental data is not the only possible. Iterative methods like presented e.g. in Refs. \cite{35gravel,36iterac} could, probably work, however, the least squares fit assures a convenient estimation of statistical errors of the extracted distributions.
The presented method of the in-situ measurements of the UCN velocity spectrum is a tool that can be used in many UCN precision experiments, where precise studies are carried out using UCN and the knowledge of the energy spectrum or changes of its shape over time has a significant impact on the estimation of systematic errors. 

\begin{acknowledgments}
We would like to thank the technical staff of Jagiellonian University and Paul Scherrer Institute for their support during the construction of the spectrometer and measurements. We acknowledge the UCN source operation group BSQ for their valuable support. 
This work has been supported by dedicated funding from National Science Centre, Poland, grant No. UMO-2016/23/D/ST2/00715 and UMO-2020/37/B/ST2/02349.
\end{acknowledgments}

\end{document}